# Mechanical Switching of Nanoscale Multiferroic Phase Boundaries

Y. J. Li[1], J. J. Wang[2], J. C. Ye[3], X. Ke[4], G. Y. Gou[5], Y. Wei[6], F. Xue[7], J. Wang[1], C. S. Wang[1], R. Peng[2], X.L Deng[6], Y. Yang[8], X. Ren[5], L-Q. Chen[2,7], C-W. Nan[2] and J. X. Zhang[1*]

1, Department of Physics, Beijing Normal University, Beijing, 100875, China.
2, School of Materials Science and Engineering, Tsinghua University, Beijing, 100084, China.
3, Physical and Life Science Directorate, Lawrence Livermore National Laboratory, Livermore, CA 94550, USA.
4, EMAT (Electron Microscopy for Materials Science), University of Antwerp, Groenenborgerlaan 171, Antwerp 2020, Belgium.
5, Frontier Institute of Science and Technology, and State Key Laboratory for Mechanical Behavior of Materials, Xi'an Jiaotong University, Xi'an 710049, China..
6, Department of Geriatric Dentistry, Peking University School and Hospital of Stomatology, Beijing, 100081, China.
7, Department of Materials Science and Engineering, Pennsylvania State University, University Park, PA 16802, USA.
8, Centre for Advanced Structural Materials and Department of Mechanical and Biomedical Engineering, City University of Hong Kong, Tat Chee Avenue, Kowloon Tong, Kowloon , Hong Kong , China.

Email: jxzhang@bnu.edu.cn

Abstract

Tuning the lattice degree of freedom in nanoscale functional crystals is critical to exploit the emerging functionalities such as piezoelectricity, shape-memory effect or piezomagnetism, which are attributed to the intrinsic lattice-polar or lattice-spin coupling. Here we report that a mechanical probe can be a dynamic tool to switch the ferroic orders at the nanoscale multiferroic phase boundaries in $BiFeO_3$ with a phase mixture, where the material can be reversibly transformed between the "soft" tetragonal-like and the "hard" rhombohedral-like structures. The microscopic origin of the non-volatile mechanical switching of the multiferroic phase boundaries, coupled with a reversible 180° rotation of the in-plane ferroelectric polarization, is the nanoscale pressure-induced elastic deformation and re-construction of the spontaneous strain gradient across the multiferroic phase boundaries. The reversible control of the room-temperature multiple ferroic orders using a pure mechanical stimulus may bring us a new pathway to achieve the potential energy conversion and sensing applications.

**Introduction**

Reversible structural transformations in crystalline solids, associated with the abrupt change in the lattice degree of freedom, give rise to a multitude of emerging phenomena and functionalities, such as large piezoelectricity across a morphotropic phase boundary [1] or a triple point involving in a metal-insulator transition of metallic vanadium oxide [2], etc. Those structural transformations in correlated systems can, therefore, create a pathway to achieve the coupling between the lattice order and other orders such as spin, charge, and orbital, where the derived functionalities above room temperature provide great opportunities for potential applications such as electromechanical or magnetoelectronic nanodevices. Multiferroic materials can provide the multifunctionalities because of the co-existence of multiple ferroic orders such as ferroelectricity, ferroelasticity, or ferromagnetism [3-8]. With a high ferroelectric and antiferromagnetic transition temperature (i.e., $T_C$=830 °C and $T_N$=370 °C) [9], Bismuth ferrite (BFO) with a rhombohedral-like (R) structure has attracted enormous attention in the past decade because of its room-temperature coupling between ferroelectricity and antiferromagnetism [10, 11]. Benefitting from the theoretical work [12] and advanced crystal growth techniques [13], it was discovered that a large compressive strain could help to stabilize a so-called tetragonal-like (T) BFO with a large tetragonality and ferroelectric polarization [14]. A self-assembled mixture of R- and T-BFO has been also stabilized by partially releasing its epitaxial strain, providing a brand-new platform to explore its structure and ferroic orders near the nanoscale multiferroic phase boundaries (~4-5 nm) [15]. Phenomena such as large flexoelectricity [15, 16] and piezoelectricity [17], shape-memory effect [18], electronic conduction [19, 20] and enhanced magnetism [21] have been studied in the nanoscale two-phase mixture, indicating the co-existence and strong coupling of multiple ferroic orders (ferroelectric, ferroelastic and magnetism) above room temperature. Emergent multiferroicity in mixed-phase BFO has triggered the active exploration of deterministic control of the multi-functional boundaries and their variants by using electric fields [22-25], temperature [26], or chemical doping [19, 20] etc.

Very recently, it is shown that the stress gradient may play a critical role on the formation and switching of ferroelectric polarization [27]. The effective flexoelectric field from the strain gradient in a polar structure is regarded as the driving force to skew the double wells which are otherwise degenerate in its free-energy profile and may flip the polarization [28, 29]. A pure mechanical force can therefore be used as a dynamic tool so that ferroelectric memory bits may be written mechanically and read electrically [30-32]. However, in a ferroelectric thin film with a uniform strain, a reversibly mechanical switching of the polarization seems to be difficult. BFO thin films with an R/T phase mixture offer an exciting opportunity for engineering the large spontaneous strain gradients at nanoscale. There undergoes an iso-symmetric phase transformation around the boundary [33, 34] with a perpendicular spontaneous local strain gradient in excess of $10^7$ m$^{-1}$ [, 15]. Additionally, an earlier study demonstrated that the application of a quasi-uniform mechanical pressure on BFO with the R/T mixture results in a structural transformation from the mixed-phase to a pure R phase accompanied by a reversible strain of ~5% [17], indicating a strong interaction between polarization and elastic orders in the nanoscale phase mixtures. Therefore a natural question arises: can a nanoscale mechanical force be used as a control parameter to reversibly re-construct the spontaneous strain gradient and the ferroic orders at the multiferroic phase boundaries?

In this work, we demonstrate that the nanoscale phase variants can be re-oriented by using a nanoscale mechanical probe, switching of such variants being usually controlled by electrical stimuli. The nanoscale multiferroic systems can be switched between "soft" (T) and "hard" (R) structures, fully characterized by the static and dynamic mechanical behaviors of the multiferroic phase boundaries. The mechanical control of the phase variants is attributed to the probe-pressure-induced local elastic deformation and the consequent re-construction of the spontaneous strain gradient and flexoelectric field among the stripe-like phase mixture. The reversible switching of the multiple ferroic orders (e.g., 180° rotation of the in-plane ferroelectric polarization, ferroelasticity), using a pure mechanical stimulus, provides an additional opportunity to study the coupling between crystal lattice and other order parameters such as spin, orbital and charge etc. This study also provides a pathway to achieve the mechanical-controlled multiple ferroic orders at room temperature.

**Results and Discussion**

**Mechanical modulus across the multiferroic phase boundaries.** A multiferroic model system of BFO thin films (~130 nm) with an R/T phase mixture have been synthesized on (001)-oriented LaAlO$_3$ (LAO) substrates using pulsed-laser deposition. In order to characterize the ferroelectric domain, epitaxial (La,Sr)CoO$_3$ (~10 nm thick) was grown on LAO before the deposition of BFO. A typical stripe-like phase mixture can be seen in the Atomic Force Microscope (AFM) image in Fig. 1 (a). The detailed growth condition can be found in the Methods Section. The nanoscale mechanical behaviors have been studied using Quantitative Nanomechanical Property Mapping (QNM) technique based on the AFM setup with a spatial resolution better than 10 nm [35-37]. The effective modulus is calculated by extrapolating the retraction curve close to the contact point and using a Derjaguin–Muller–Toporov (DMT) model [38] as shown in supplementary Fig. S1 (the green line). The details can be found in the Methods Section. A mechanical force (~5 µN) was applied during the measurement, which is much lower than the threshold value of the structural deformation of the mixed-phase BFO [39]. The effective Young's modulus ($E$) has been mapped across the multiferroic phase boundaries with a high spatial resolution. As seen in Fig. 1 (b), the mapping of the effective modulus of the mixed-phase BFO was also observed as stripe-like features, which gives us a first glimpse of the inhomogeneous effective modulus across the multiferroic phase boundaries. The full understanding of the effective modulus and the critical pressure-induced elastic deformation in the mixed-phase BFO is the basis to achieve the mechanical control of the multiferroicity at nanoscale.

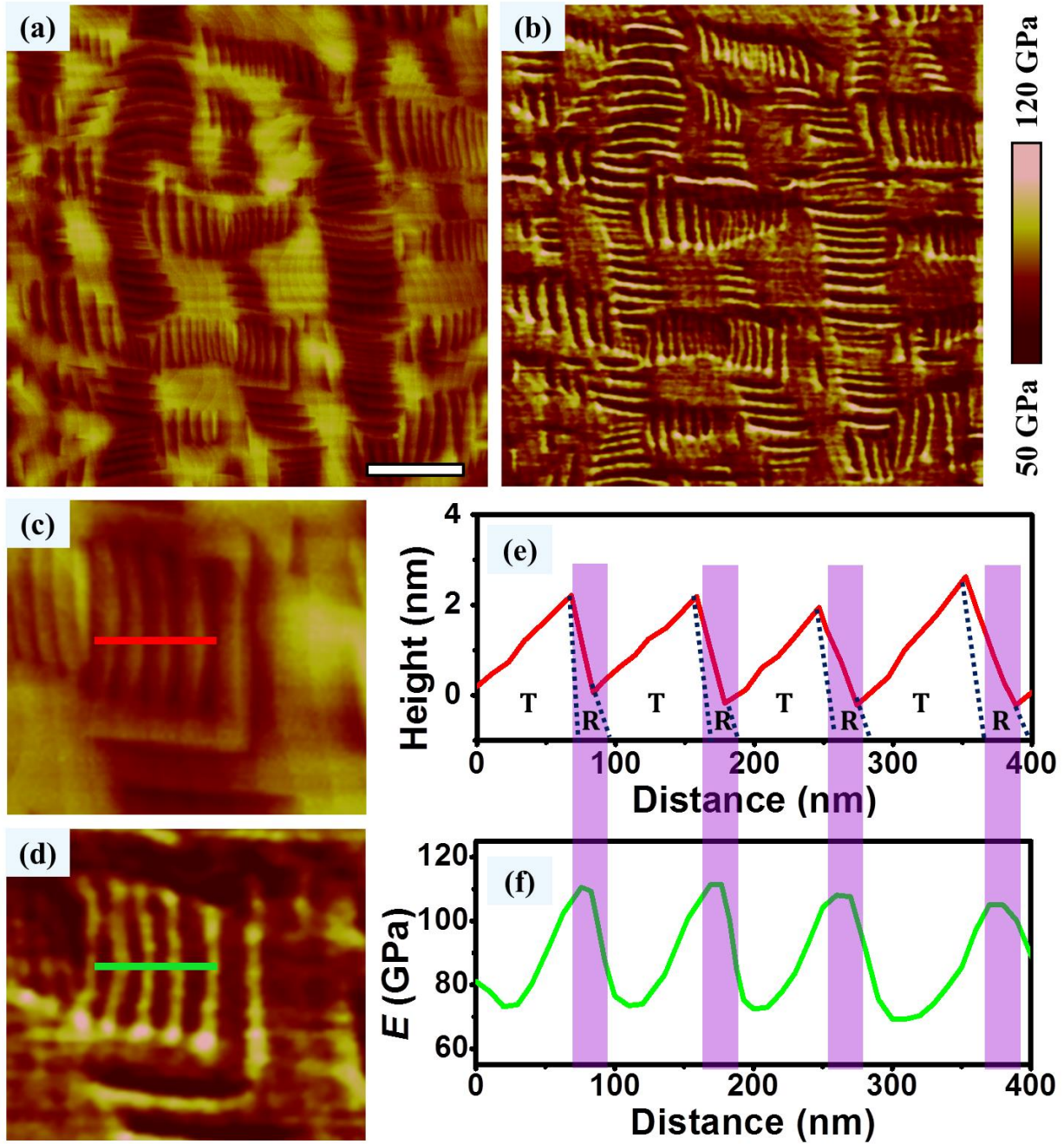

Fig. 1. (a) Topography of a typical mixed-phase BFO film with stripe-like features. (b) Corresponding effective *E* mapping obtained by AFM-based QNM. High-resolution topography (c) and effective *E* (d) across the multiferroic phase boundaries. Line profiles of the surface

topography (e) and effective $E$ (f), where distinct effective $E$ were observed between T and R phases. The scale bar is 1 µm.

A close look at the mechanical behaviors at the boundaries using the QNM method further indicates that there is a remarkable inhomogeneity of the effective modulus between the R and T phases, as seen in Fig. 1 (c) and (d). Line profiles of the atomic surface topography (red) and effective modulus (green) across the R/T boundaries are shown in Fig. 1 (e) and (f) respectively. We noticed that the effective modulus changes from ~75 GPa to ~110 GPa across the R/T mixture, where the maximum and minimum values were observed to occur when the R and T phases are exposed to the film surface respectively. It is known that the mechanical behaviors of a thin film measured via the QNM method are primarily from the near-surface contributions [33, 34]. This result, therefore, possibly indicates that the effective modulus in T phase is much smaller than that in R phase. The distinct difference of effective modulus between R and T phases has been further studied on respective pure phases in the matrix of the mixed-phase BFO. For pure T- and R-phase BFO among the phase mixture, the effective moduli were measured as ~70 GPa and ~140 GPa, providing direct evidence that the T phase is much "softer" than R phase in mixed-phase BFO, as seen in supplementary Figs. S2. In addition, the effective modulus of the pure R phase is also well consistent with its bulk value [40, 41]. The detailed formation of the pure T phase (by electrical poling) and R phase (by mechanical indentation) in the mixed-phase matrix can be seen in supplementary information.

**Dynamic measurements of the effective modulus and mechanical behavior during phase transformation.** In order to obtain the threshold pressure of the structural transformation across the multiferroic phase boundaries, dynamic measurements of the effective modulus have been carried out using nanoindentation [42, 43]. A Berkovich indenter (Hysitron nanoindentation system) with a spherical tip was employed on top of the pure T phase among the mixed-phase BFO matrix. The detailed measurements can be seen in the Methods Section. The typical load-depth curve with a maximum load of 1 mN is shown in Fig. 2 (a), where we can observe three characteristic periods. With the increase in loading force, T phase initially deforms (period I) and then gradually transforms into R phase with an indentation depth of 18±0.2 nm, corresponding to a phase-transformation-induced (T to R) strain of ~14% [19]. This period can be fitted by the Hertz equation [44]. The effective modulus during the phase transformation is 56±13 GPa. A further increase of the loading force deforms the pure R-phase crystal elastically in period II, which can be fitted by a modified equation [45]. The nature of the elastic deformation after the phase transformation could be proved through the reversible load response in region II, as shown in supplementary Fig. S3. The derived effective modulus during the elastic deformation of pure R phase is ~169±11 GPa. In period III, when the load approached the yield stress for R phase, the BFO film fractures with a "pop-in" behavior, indicating an abrupt plastic damage of the pure R phase under a large loading force. In order to simultaneously monitor the phase transformation induced by the nanoindentation, topographical images corresponding to each period in the characteristic load-depth curves are shown in supplementary Fig. S4. They further demonstrate the deformation sequences of the T/R transformation, elastic R and pop-in plastic damage. Furthermore, load-depth curves were also performed using the AFM-based nanoindentation technique [46] as shown in Fig. 2 (b) and Fig. 2 (c). 3000 load-depth curves were collected from T and R phases in the mixed-phase BFO. Statistical distributions demonstrate that the effective modulus of the T and R phases are about 60 GPa (Fig. 2 (b)) and 150 GPa (Fig. 2 (c)) respectively, as determined by the peaks with a Gauss fittings. These values obtained from the AFM-based nanoindentation agree well with the values obtained from the static QNM measurement. From the study of the dynamic indentation in mixed-phase BFO, it is therefore clear that the threshold probe pressure to induce a structural transformation is 5.7±1.2 GPa and the pop-in plastic damage threshold is 14.4±1.2 GPa. The detailed fitting can be seen in supplementary information.

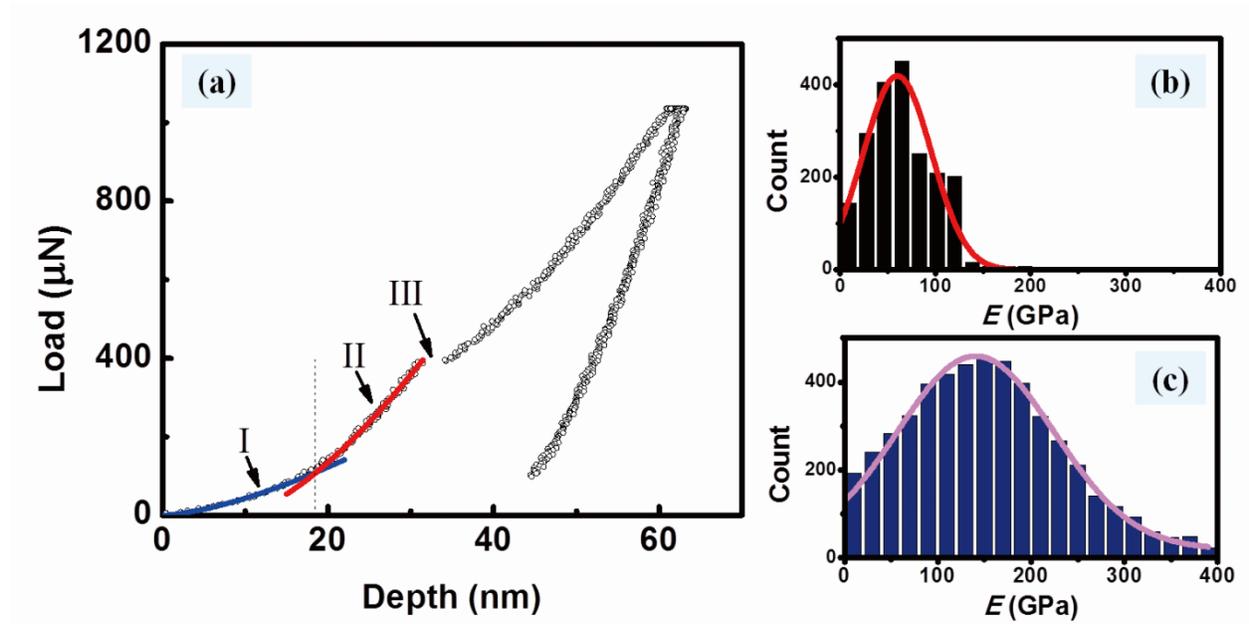

**Fig. 2.** (a) Typical load-depth curve of BFO with a phase mixture from regular nanoindentor, showing dynamic behavior of the phase transition of BFO. The Young's modulus for T- and R-phase BFO are 56±13 GPa (blue line fitting) and 169±11 GPa (red line fitting), respectively. (b) and (c) Statistical distributions of T-phase and R-phase Young's modulus of BFO, fitting from 3000 load-depth curves obtained from AFM-based nanoindentation with effective $E$ for T-phase BFO of ~60 GPa and R-phase BFO of ~150 GPa, respectively.

**Switching of multiferroic phase boundaries by a local scanning probe pressure.** With the full understanding of the effective modulus and the threshold pressure of local phase transformations in mixed-phase BFO, we now turn to the nanoscale control of the ferroic orders (e.g. ferroelastic or ferroelectric) at the boundaries, where we can observe that the projections of in-plane ferroelectric polarizations are opposite near boundaries with different tilting directions as shown in Fig. S5. A local mechanical stress of 9.5±2.3 GPa (lower than the threshold value of ~14.4±1.2 GPa for the plastic damage) was applied on the nanoscale uni-directionally scanning probe with a contact mode. The detailed measurement of piezoresponse force microscopy (PFM) and the mechanical control process can be seen in the Methods Section, supplementary information and Fig. S6, respectively. Fig. 3 shows the topographical AFM images (top), in-plane ferroelectric domains (middle) and Young's modulus (bottom) before (a) and after (b) the nanoscale probe-induced mechanical switching. It shows two structural variants (defined by the tilting angle (α and β) of the boundaries) of stripe-like features before and after the application of a scanning probe stress at the same location of the mixed-phase BFO. These mechanical switching indicated that the projection of the in-plane ferroelectric polarization underwent a 180° rotation across the multiferroic phase boundaries, accompanying a distinct change of elastic modulus. Different from the conventional electrical tuning of the phase variants [23], this observation demonstrates that the nanoscale multiferroic phase structures can be completely switched through a pure mechanical technique at room temperature. The nanoscale non-volatile switching of the multiferroic phase boundaries using a mechanical probe as a dynamic tool is schematically described in Fig. 3 (c).

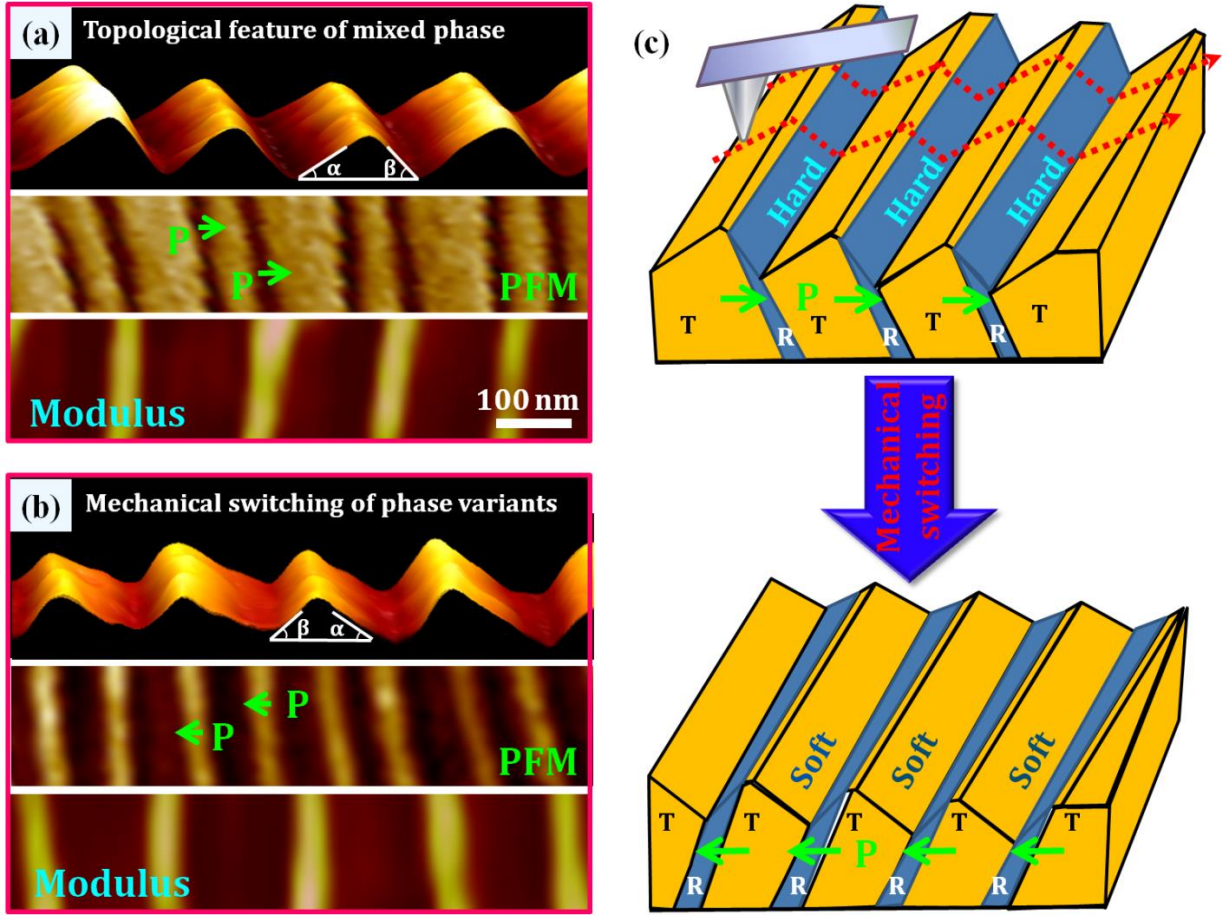

**Fig. 3.** Three-dimensional AFM images (top), in-plane PFM images (middle) and effective modulus images (bottom) of stripe-like BFO with a phase mixture before (a) and after (b) the probe scanned perpendicular to the boundaries with a pressure of ~9 GPa. Green arrows describe the projections of the in-plane polarization near the multiferroic phase boundaries. (c) Corresponding schematic diagrams illustrate the scanning probe control process of the re-orientation of the mutiferroic phase boundaries. The dashed arrows in the schematics describe the trace of the scanning probe with a mechanical pressure applied on mixed-phase BFO.

**Microscopic structure of the multiferroic phase boundaries.** In polar structures, the application of an inhomogeneous force can either generate a strain gradient in materials with a uniform polarization [27] or re-construct the strain gradient due to elastic deformation in structures with spontaneous flexoelectric orders such as domain walls [8] or phase boundaries [15]. The multiferroic phase boundaries in the mixed-phase BFO thin film provide us a good platform to probe their nanoscale elastic deformation and flexoelectricity resulting from the intrinsic coupling of strain gradient and polarization. A pure mechanical deformation is not sufficient to re-orient the phase variants which can be seen in our previous studies [17] and also supported by the phase-field simulations, where the phase morphology mostly remains without considering the flexoelectricity at the boundaries (detailed simulation can be seen in supplementary information and Fig. S7). This means that the spontaneous flexoelectricity across the multiferroic phase boundaries would be critical to the control of non-volatile phase boundaries. As seen in Fig. 4 (a), typical orders of inclined boundaries in mixed-phase BFO are shown in the cross-sectional scanning transmission electron microscopy (STEM) image, acquired using high-angle annular dark field STEM (HAADF-STEM). The presence of R-BFO, T-BFO and the LAO substrate is confirmed from the Fast Fourier Transform (FFT) pattern as seen in the inset. The conditions for sample preparation and imaging are shown in the Methods Section. The detailed atomic structures across the multiferroic phase boundaries show that the lattice gradually transits between R and T as the epitaxial-strain-driven morphotropic or iso-structural phases [13]. Obviously, they have been categorized into two types of boundaries (R/T and T/R boundaries defined by the in-plane polarization direction). Based on high-resolution HAADF-STEM images, strain gradients across the R/T and T/R boundaries were evaluated using geometrical phase analysis (GPA), where the strain $\varepsilon_{xx}$ was mapped using the non-deformed LAO substrate as reference. Detailed measurement could be found in the Methods Section. The maximum in-plane strain gradients at the R/T and T/R boundaries are $-7.7 \times 10^6$ m$^{-1}$ and $7.1 \times 10^6$ m$^{-1}$, respectively (see Figs. 4 (b) and 4 (c)). In polar structures with the strain gradient and flexoelectricity [47-50], the effective flexoelectric field [51,52] can be expressed as:

$$E_S = \frac{e}{4\pi\varepsilon_0 a}\frac{\partial u}{\partial x}, \qquad (1)$$

where $e$ is the electronic charge, $\varepsilon_0$ is the permittivity of free space, $a$ is the in-plane lattice constant of R-phase BFO, and $\partial u/\partial x$ is the in-plane strain gradient near the boundary. The estimated maximum flexoelectric fields are about $-0.29\ MV/cm$ ($E_{S(R/T)}$) and $0.27\ MV/cm$ ($E_{S(T/R)}$) for R/T and T/R boundaries respectively. The total elastic and electrostatic potentials reach minima with such a flexoelectric orders. Namely, the spontaneous in-plane flexoelectric fields cancelled each other out in the film plane (fields in both boundaries are anti-parallel and perpendicular to the as-grown stripe-like orders), as described by the blue and brown arrows in Fig. 4 (a).

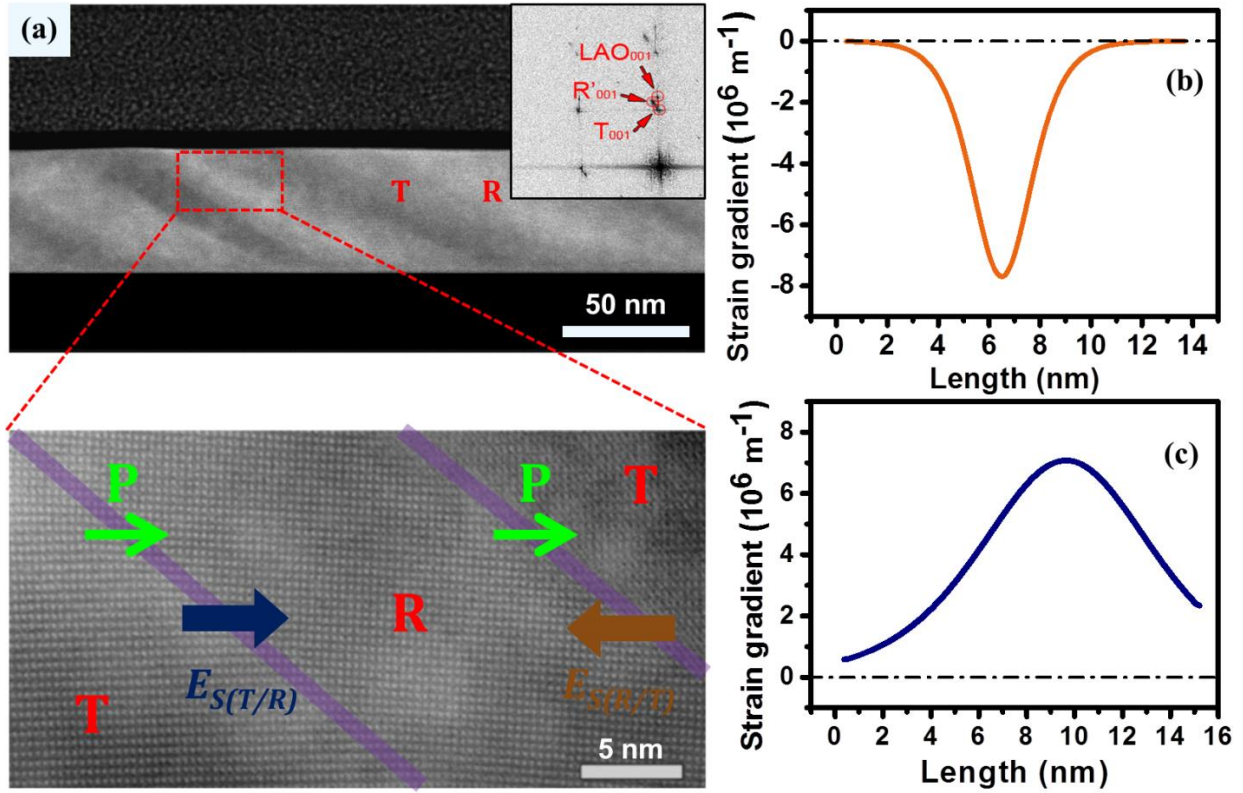

**Fig. 4.** (a) Cross-sectional HAADF-STEM image of the mix-phase BFO, viewed along mix-phase BFO [010]. Typical orders of inclined boundaries are presented. The inset shows the FFT of mixed-phase thin film on LAO substrate. A high-resolution image is at the bottom, showing inner flexoelectric fields are reversed in R/T and T/R boundaries described with brown and blue arrows. The in-plane projections of the ferroelectric polarization are indicated by the green arrows. (b-c) The in-plane strain gradients across the R/T and T/R boundaries obtained from the GPA analysis.

When a local scanning pressure of up to 9.5±2.3 GPa (larger than the value of phase transformation and smaller than the one of pop-in threshold) was applied on top of the phase mixture, the boundaries underneath the force probe (contact radius less than 20 nm) can be elastically erased and consequently break the initial flexoelectric fields across the boundaries. The local remaining flexoelectric field $E_{S(R/T)}$ (brown arrow) or $E_{S(T/R)}$ (blue arrow) will be applied on the local boundary in the film plane. As seen in the PFM and STEM images, projections of in-plane polarization near R/T and T/R boundaries are anti-parallel to the $E_{S(R/T)}$ and parallel to $E_{S(T/R)}$. In order to accomplish the coupling between polarization and flexoelectric fields so that the switching of the multiferroic phase boundaries can be achieved, the anti-parallel field to the polarization contributes to the polar rotation and switching, but the parallel field to the polarization does not. Consequently, erasing of the T/R boundary due to the elastic deformation and the remnant local in-plane net field ($E_{S(R/T)}$) induces the coupling between polarization and re-constructed strain gradient, but mechanical erasing of R/T boundary and the remnant $E_{S(T/R)}$ does not. Therefore, during the application of a local scanning probe pressure perpendicular to the multiferroic phase boundaries, the re-construction of local strain gradients and the $E_{S(R/T)}$ across the boundaries re-orients the multiferroic phase variants, regardless of the probe scanning from R/T to T/R or from T/R to R/T as seen in supplementary Fig. S8. It is worthwhile to mention that the effective flexoelectric field perpendicular to the boundaries in the film plane is well above the critical electrical switching field ($< 0.25\ MV/cm$) along the same direction as reported elsewhere [23], further implying that the mechanical switching of the phase variants are plausible from the electromechanical coupling point of view.

**Local mechanical pressure on T/R or R/T boundaries respectively.** In order to provide direct evidence of the above scenario, a local pressure was applied specifically on the T/R or R/T boundaries by programming the probe scanning direction parallel to the multiferroic phase boundaries. The detailed control process has been programmed as seen in supplementary Fig. S9. When the local pressure (~9 GPa) was only applied on the T/R boundaries (Fig. 5 (a)), we observe that re-orientation of the multiferroic phase boundaries occurs before (Fig. 5 (b)) and after (Fig. 5 (c)) the mechanical scanning as shown in the AFM images and the corresponding line profiles. However, when the local pressure was only applied on the R/T boundaries (Fig. 5 (d)), we observe no change of the phase mixture before (Fig. 5 (e)) and after (Fig. 5 (f)) the mechanical scanning until the pop-in plastic damage. In Fig. 5 (a) and (d), the dashed red arrows schematically illustrate the trace of the mechanical scanning probe selectively applied on the T/R or R/T boundaries. It is therefore demonstrated that only the suppression of the T/R boundaries can give rise to the strong coupling of polarization and flexoelectricity, so that the switching can be achieved by a local mechanical probe.

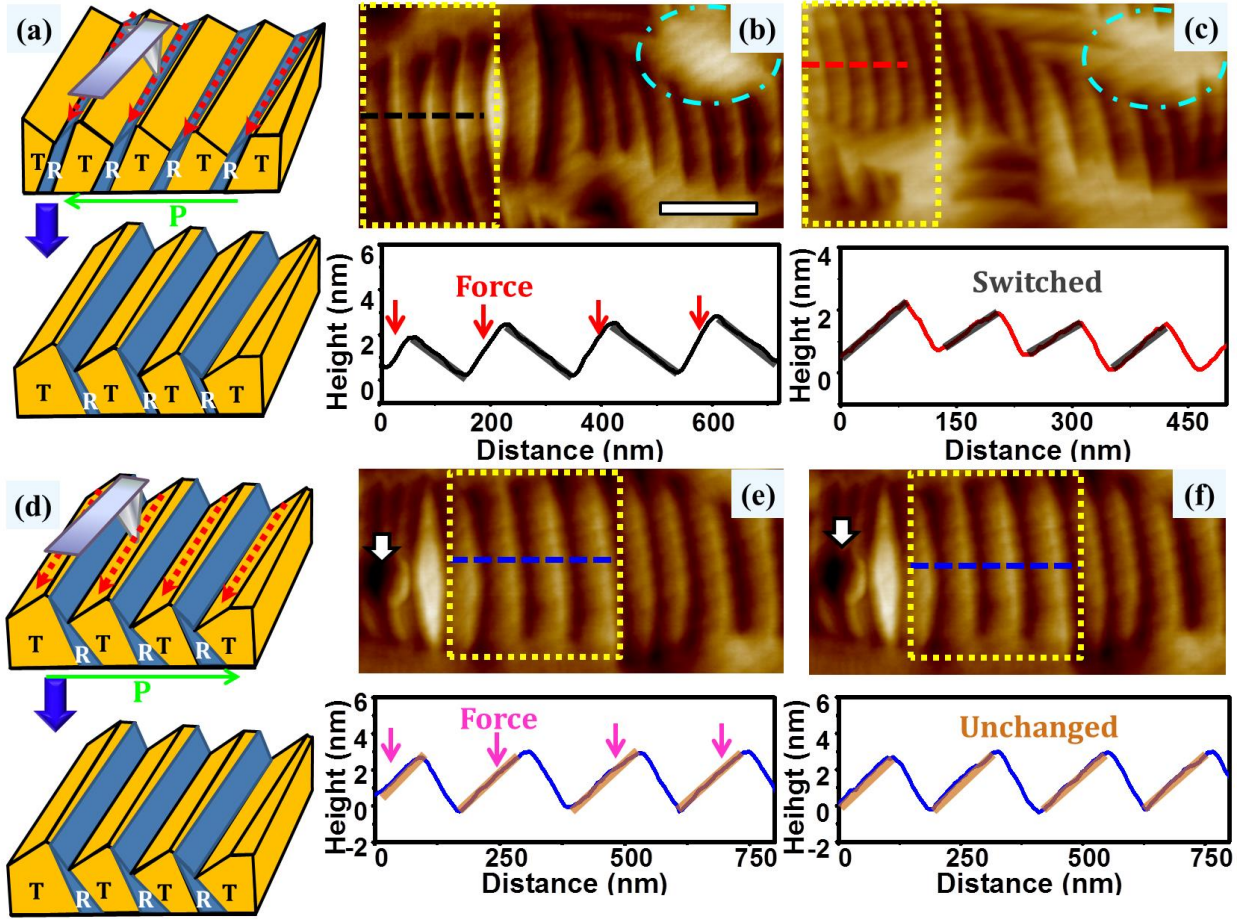

**Fig. 5.** (a) Schematic diagram of the local mechanical pressure applied on T/R boundaries when the probe scanned along parallel to the multiferroic phase boundaries. AFM images and the corresponding line profiles before (b) and after (c) the applications of the scanning probe pressure, showing the re-orientation of the ferroelectric boundaries. (d) Schematic diagram of the local mechanical pressure applied on R/T boundaries when the probe scanned along parallel to the multiferroic phase boundaries. AFM images and the corresponding line profiles before (e) and after (f) the applications of the scanning probe pressure, demonstrating that no change occurs. The scale bar is 200 nm. The dashed ellipse and white arrows in (b), (c) and (e), (f) indicate the markers during the mechanical control.

**Conclusion**

The nano-mechanical properties including effective modulus of mixed-phase BFO with an ultrahigh electromechanical coupling were investigated using QNM technique and conventional nanoindention. With the understanding of the local pressure-induced phase transformations across the multiferroic phase boundaries, the phase variants and orientations of the multiferroic phase boundaries can be reversibly controlled via a pure mechanical stimulus. Detailed studies reveal that the elastic deformation and the consequent re-construction of the as-grown flexoelectric fields are the nanoscale origin of the non-volatile structural control. Apart from the mechanical modulation of the 180° rotation of the in-plane ferroelectric polarization, this work further demonstrates that, at these multiferroic phase boundaries [17,18,21], a mechanical force may also be a possible dynamic tool to control the lattice-polar or even lattice-spin coupling at room temperature, which may open a new route to design electromechanical, data-storage and energy-conversion applications at nanoscale.

## 4  Methods Section

**4.1 Thin film growth.** Mixed-phase BFO thin films were grown on (001)-oriented LAO substrates using pulsed-laser deposition. The growth temperature was maintained at 700 °C, and the working oxygen pressure was 12 Pa. For all films, a growth rate of 2 nm/min and a cooling rate of 5 °C/min under $10^4$ Pa were used. A laser energy density of 1.2 J/cm$^2$ and a repetition rate of 5 Hz were used during the deposition. The thickness of the mixed-phase films was controlled at ~130 nm.

**4.2. Characterization of the QNM.** QNM is an AFM-based technique (Bruker, Multimode 8) to study the mechanical properties across a surface with an ultrahigh spatial resolution [37]. A diamond tip with spring constant of ~213 N/m was used for QNM imaging. The effective modulus is calculated by extrapolating the retraction curve close to the contact point and using a Derjaguin–Muller–Toporov (DMT) model [38] (equation 1) as shown in supplementary Fig. S1 (the green line):

$$F_{tip} - F_{adh} = \frac{4}{3}E^*\sqrt{Rd^3} \qquad (1)$$

where $F_{tip}$ is the loading force, $F_{adh}$ is the adhesion force, $E^*$ is the reduced Young's modulus, $R$ is the tip radius and $d$ is the deformation depth. And the $E$ can be calculated by $E^*$ and Poisson's ratio of sample and tip. The details can be seen in supplementary information. For measurement of the effective modulus of mixed-phase BFO, the indentation depth is controlled less than 5 nm, around 4% of the thin film thickness (~130 nm). Therefore, the contribution from the substrate on the mechanical modulus can be negligible.

**4.3 Nanoindentation.** Spherical indentation was conducted on the top surface of the T-phase BFO using the Hysitron™ TriboIndenter (Hysitron, MN, USA) equipped with a modified Berkovich indenter. The tip radius $R=275 \pm 26$ nm was calibrated using a standard fused silica sample ($E^*$=69.6 GPa). The $E^*$ can be extracted by the Hertz equation (2) [44] or the modified equation (3) [45]:

$$P = \frac{4}{3}E^*R^{1/2}h_e^{3/2} \qquad (2)$$

where $P$ is load, $h_e$ is depth, and $R$ is tip radius.

$$P = \frac{4}{3}E^*R^{1/2}(h_e - h_0)^{3/2} \qquad (3)$$

where $P$ is load, $h_e$ is depth, $R$ is tip radius and $h_0$ is the effective zero point of the contact of the R phase. The detailed fitting can be seen in supplementary information.

AFM-based nanoindentation was carried out on mixed-phase BFO using the same diamond tip as the one in QNM experiments with ~5 µN loads force and three-second indentation times. The effective modulus, $E^*$, was calculated from the unloading curve. ~3000 load-depth curves were collected.

**4.4 PFM measurement across the phase boundaries.** PFM measurements were carried out on a Bruker Multimode 8 AFM with commercially available TiPt-coated Si tips with spring constant of ~5 N/m (Mikro Masch). The typical tip velocity was 2 µm/s, and the amplitude and frequency of the AC input were 0.5 $V_{pp}$ and 22 kHz respectively. The poling voltage on the electrically biased tip was controlled up to 20 V. High-resolution PFM measurements were completed on a wide array of samples. The polarization vectors near the boundaries have been re-constructed based on the domain images obtained from the cantilever scanning parallel and perpendicular to the boundaries.

**4.5 TEM sample preparation and imaging.** A thin cross-sectional lamella for TEM study was prepared by Focused Ion Beam (FIB) using a FEI-Helios NanoLab DualBeam system. A protective layer of amorphous carbon and Pt layers were deposited on the sample surface prior to milling. The lamella was prepared from the area where typical stripe patterns were presented. HAADF-STEM images were acquired on a FEI Titan Cube microscope fitted with a corrector, operated at 300 kV using imaging condition as reported before [18]. Strain gradients across the phase boundaries were mapped out using GPA to evaluate the strain gradient across the T/R boundary and R/T boundaries. [100]pc and [001]pc directions (i.e. in-plane vector and out-of-plane vector) are used for calculating the phase maps, which are the x and y directions respectively. X direction is along the fast scanning direction and thus the obtained strain ($\varepsilon_{xx}$) is valid in calculating the strain gradient. LAO substrate away from the interface is used as reference.

## Acknowledgements

This work was supported by the National Science Foundation of China under contract Nos. 51322207, 51332001, 11274045 and the National Basic Research Program of China, under contract No. 2014CB920902. The authors also thank Dr. Hao Sun and Dr. Dengli Qiu of Bruker BNS China for their kindly discussion and support about QNM. L-Q. Chen was supported by the U.S. Department of Energy, Office of Basic Energy Sciences, Division of Materials Sciences and Engineering under Award FG02-07ER46417 and National Science Foundation of China (No. 51472140). The work from Xi'an Jiaotong University is supported by National Basic Research Program of China (No.2012CB619402), National Science Foundation of China (No.11204230),